\documentclass[reprint, showpacs]{revtex4-1}
\usepackage{graphicx}
\usepackage{amsfonts}
\usepackage{natbib}

\begin{document}
\title{Quantum localization of chaotic eigenstates and the level spacing distribution}

\author{Benjamin Batisti\'c}
\author{Marko Robnik}

\affiliation{CAMTP - Center for Applied Mathematics and Theoretical
Physics, University of Maribor, Krekova 2, SI-2000 Maribor, Slovenia, European Union}

\date{\today}

\begin{abstract}
The phenomenon of quantum localization in classically chaotic eigenstates
is one of the main issues in quantum chaos (or wave chaos), and thus plays
an important role in general quantum mechanics or even in general wave mechanics. 
In this work we propose two different localization measures characterizing 
the degree of quantum localization, and study their relation to another
fundamental aspect of quantum chaos, namely the (energy) spectral statistics.
Our approach and method is quite general, and we apply it to billiard systems. 
One of the signatures of the localization of chaotic eigenstates 
is a fractional power-law repulsion between the nearest energy levels in the
sense that the probability density to find successive levels on a distance
$S$ goes like $\propto S^\beta$ for small $S$, where $0\leq\beta\leq1$,
and $\beta=1$ corresponds to completely extended states. We show that there is a 
clear functional relation between the exponent $\beta$ and the two
different localization measures.
One is based on the information entropy and the other one on the correlation
properties of the Husimi functions. We show that the two definitions are
surprisingly linearly  equivalent. The approach is applied in the case of a
mixed-type billiard system (Robnik 1983), in which the separation 
of regular and chaotic eigenstates is performed.
\end{abstract}


\maketitle

\section{Introduction}

Quantum chaos (or more generally, wave chaos) is the study of the
signatures of classical chaos in quantum (or general wave) systems 
\cite{Stoe,Haake}. The quantum localization of classical chaotic
diffusion in the time-dependent domain is one of the most important fundamental
phenomena in quantum chaos, discovered and studied in the quantum 
kicked rotator \cite{Cas1979,Chi1981,Chi1988,Izr1990} by Chirikov, Casati,
Izrailev, Shepelyansky, Guarneri, and further developed by many others.
It was mainly Izrailev who has studied the relation between the
spectral fluctuation properties of the quasienergies (eigenphases)
of the quantum kicked rotator and the localization properties
\cite{Izr1988,Izr1989,Izr1990}. This picture is typical for
chaotic time-periodic (Floquet) systems.

In the time-independent domain the phenomenon also manifests itself
as the localization of the Wigner functions of chaotic eigenstates
in the phase space, meaning that the chaotic quantum eigenstate does not
occupy the entire available classical chaotic phase space, but is
localized on it. This aspect is very closely related, or almost equivalent,
to the Anderson localization, as shown for the first time by Fishman,
Grempel and Prange \cite{FGP1982} in the case of the quantum kicked rotator 
and studied very intensely by many others (for further references see \cite{Stoe,Haake}).
The quantum localization in billiards is reviewed by Prosen in reference 
\cite{Pro2000}.

However, one of the open questions is to define an appropriate 
measure of  localization in general, which is the main point and  
result of this paper. For this purpose we shall use the Husimi
functions, which - unlike the Wigner functions - are positive definite, 
and in fact are Gaussian smoothed Wigner functions.

Another fundamental phenomenon in quantum chaos in the time-independent
domain is the statistics of the fluctuations in the energy spectra. In analogy
with the time-periodic systems we find the relationship between
the localization measure and the spectral (energy) level
repulsion parameter, to be precisely defined below. This finding
is the main result of this work.

The statistical properties of energy spectra of quantum systems are
remarkably universal \cite{Stoe,Haake,Mehta,GMW,Rob1998}.
In a sufficiently deep semiclassical limit
they are determined solely by the type of classical motion, which can be either
regular or chaotic \cite{Percival1973,BerRob1984,Rob1998,BatRob2010,
BatRob2013}.
The classification regular-chaotic can be done by analyzing the
structure of eigenstates in the quantum phase space, based on the
Wigner functions, or Husimi functions.
The level statistics is Poissonian if the underlying classical invariant
component is regular, whilst for chaotic extended states
the Random Matrix Theory (RMT) applies \cite{Mehta}, 
specifically the Gaussian Orthogonal Ensemble statistics
(GOE) in case of an antiunitary symmetry. This is the {\em Bohigas-Giannoni-Schmit
conjecture} \cite{Cas1980, BGS1984}, which has been proven only recently 
\cite{Sieber,Mueller1,Mueller2,Mueller3,Mueller4} using the semiclassical methods
and the periodic orbit theory developed around 1970 by Gutzwiller
(\cite{Gutzwiller1980} and the references therein), an approach initiated by 
Berry \cite{Berry1985}, well reviewed in \cite{Stoe,Haake}.

In a mixed-type  case, where the
classical regular and chaotic invariant components coexist,
the Berry-Robnik (BR) picture applies \cite{BerRob1984},
based on the Principle of the Uniform Semiclassical
Condensation (PUSC) of Wigner functions on the classical invariant 
components \cite{Percival1973, Berry1977, Rob1988, Rob1998},
which allows for conceptual separation of regular and chaotic eigenstates 
\cite{BatRob2013}. 
Assuming such separation, the  Poisson statistics for the 
regular levels, whilst the GOE for the extended chaotic levels,
plus statistical independence between them, we arrive at the
BR level spacing distribution described below.

The BR theory was confirmed in several different dynamical systems
\cite{ProRob1993a,ProRob1994a,ProRob1994b,ProRob1999,Pro1998,BatRob2010,BatRob2013,
BarBet2007}
under {\em the semiclassical condition} that all classical transport times are
shorter than the Heisenberg time $t_H=2\pi\hbar/\Delta E$, where
$\Delta E$ is the mean energy level spacing.  There are two
major quantum effects in mixed-type systems calling for generalization.
First, the tunneling effect due to the finite wavelength,
which couples eigenstates from different invariant domains, and thus
breaks the assumption of the statistical independence
\cite{Vidmar2007,BatRob2010}. As the tunneling effects vanish
exponentially with the inverse effective Planck constant,
they rapidly disappear at higher energies, and can be neglected
there, which is the case in the present work.
The second is the effect of localization of chaotic eigenstates
which sets in if the above {\em semiclassical condition} is not satisfied
and is manifested  in the statistical properties different from the RMT.
Thus the localization effects are far more persistent than the tunneling
effects.

The most important statistical measure
is the level spacing distribution $P(S)$, assuming spectral
unfolding such that $\left<S\right>=1$. For integrable
systems  $P(S)=\exp\left(-S\right)$, whilst for extended chaotic systems it
is well approximated by the Wigner distribution 
$P(S)= \frac{\pi S}{2}\exp\left(-\frac{\pi}{4}\,S^2\right)$.
The distributions differ significantly in a small $S$ regime, where there
is no level repulsion in a regular system and a linear level repulsion,
$P(S)\propto S$, in a chaotic system. Localized chaotic states exhibit 
the fractional power-law level repulsion $P(S)\propto S^\beta$, as clearly
demonstrated recently by Batisti\'c and Robnik \cite{BatRob2013}.

The localization is a pure quantum effect which appears if the Heisenberg
time $t_H$,  which is the time scale on which the 
quantum evolution follows the classical one, is smaller than the relevant 
classical transport time, such as ergodic time.
Up to the Heisenberg time the quantum system behaves as if the
evolution operator had a continuous spectrum, but at times longer
than Heisenberg time the almost periodic spectrum of the evolution
operator becomes resolved, and the interference effects set in, 
resulting in a destructive interference causing the quantum localization.
The ergodic time may be very long if the chaotic region has a complicated, but
typical KAM structure, due to the presence of the partial barriers in the form of
barely destroyed irrational tori, called cantori, 
which allow for a very slow transport only. 
The weak ($\beta<1$) level repulsion of localized states is a consequence of
the loose coupling of states which are divided by such partial barriers, but 
the whole distribution $P(S)$ is globally not known.
Several different distributions which would extrapolate the small
$S$ behaviour were proposed. The most popular are the Izrailev
distribution \cite{Izr1988,Izr1989,Izr1990} and the Brody distribution
\cite{Bro1973,Bro1981}. 
Brody distribution is a simple generalization of the Wigner distribution, 
$P(S)=c\,S^{\beta}\,\exp{\left(-d\,S^{\beta}\right)}$, where $c$ and
$d$ are the normalization constants determined by
$\left<1\right>=\left<S\right>=1$. 
It interpolates the
exponential and Wigner distribution as $\beta$ goes from $0$ to $1$. 
Izrailev distribution is a bit more complicated but has a feature that it is a
better approximation for the GOE distribution at $\beta=1$. One 
important theoretical plausibility argument of Izrailev in support of
such intermediate level spacing distributions is that
the joint level distribution of Dyson circular ensembles can be extended
to noninteger values of the exponent $\beta$ \cite{Izr1990}.
However, recent numerical results show that Brody distribution 
is slightly better in describing real data
\cite{BatRob2010,BatRob2013,ManRob2013,BatManRob2013}, 
and is simpler, which is the reason why we prefer and use it.

In the absence of the tunneling  effects between the regular and chaotic
eigenstates, the BR picture applies, but must be generalized to 
take into account possible effects of localization of chaotic eigenstates.
The BRB  distribution was proposed in
\cite{ProRob1994a,ProRob1994b,BatRob2010}.  
The difference from the original BR distribution is that the 
limiting GOE distribution for chaotic levels is
now replaced with the Brody distribution. The BRB distribution  
can be written most compactly in terms of a gap probability $E(S)$,
which is a probability that an interval of length $S$ is empty of levels
in the unfolded spectrum ($\left<S\right>=1$). Namely, the general
relation $P(S)=d^2E(S)/dS^2 $ holds.
Due to the independence of sub-spectra, 
the BRB gap probability is just a product of gap
probabilities for chaotic and regular levels and thus the BRB level
spacing distribution equals
\begin{equation}
  P_{\mathrm{BRB}}(S)=
\frac{d^2}{dS^2}\left(E_{\exp}(\rho_r\,S)\,E_{\mathrm{Brody}}(\rho_c\,S)\right)
  \label{eqDefBRB}
\end{equation}
where $\rho_r$ and $\rho_c=1-\rho_r$ are the relative classical phase space 
volumes of a regular and a chaotic domain, respectively. 
Here only the dominant chaotic component is considered, the much smaller ones are
neglected, which usually is an excellent approximation.
The BRB distribution has two parameters, one classical, $\rho_r$, and one quantal,
$\beta$. Numerical results show that BRB gives an excellent description
\cite{ProRob1994a,ProRob1994b,BatRob2010,BatRob2013}. 

\begin{figure}
  \centering
  \includegraphics{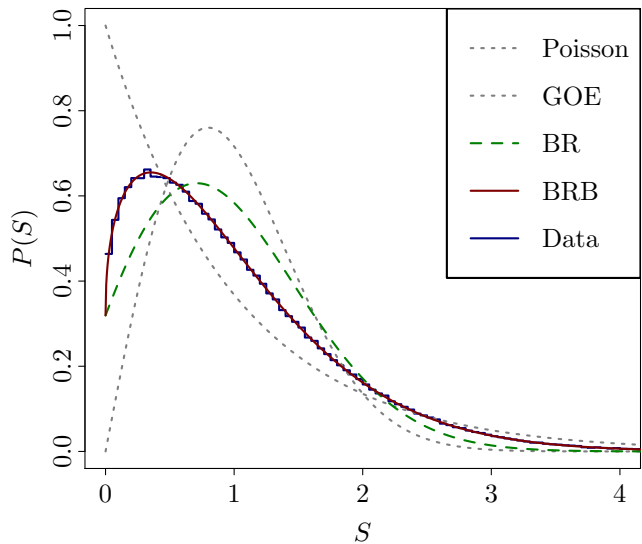}
  \caption{Nearest levels spacings distribution for $\lambda=0.15$
  billiard around $k\approx2000$. The BRB distribution fits data very
well. The classical $\rho_r=0.175$ was used and $\beta=0.45$ was obtained
from the fitting.}
  \label{figPS}
\end{figure}

The open question is how does the parameter $\beta$ depend on the
localization.
This question was raised for the first time by Izrailev
\cite{Izr1988,Izr1989,Izr1990},
where he numerically studied the quantum kicked rotator, which is a 
1D time-periodic system. His result showed that the parameter $\beta$, which
was obtained using the Izrailev distribution, is functionally related to the
localization measure defined as the information entropy of the
eigenstates in the angular momentum representation. His results were
recently confirmed and extended, with the much greater numerical accuracy
and statistical significance \cite{ManRob2013,BatManRob2013}.

In this paper we show for the first time that there is indeed a functional
relation between the level repulsion parameter $\beta$ and the localization 
measure also in autonomous quantum systems, in perfect analogy with the
quantum kicked rotator.  We define two different
general localization measures in terms of the Husimi 
functions and show that they are equivalent. Our approach is quite
general, but will be demonstrated in the case of the billiards as model systems.

\section{Localization measures}

The Wigner functions, defined in the phase space $(q,p)$, 
 have been introduced by Wigner in 1932 \cite{Wig1932}.
They are real valued, but not positive definite. Usually they oscillate
around the zero value in regions which do not have much physical significance
and obscure the picture, so one would prefer to smooth out such
fluctutations. The Husimi
functions \cite{Hus1940}, also called Husimi quasi-probability distributions,
however, are positive definite.  They are
defined as Gaussian smoothed Wigner functions, or equivalently, as a
square of the projection of the eigenfunction of the corresponding eigenstate
onto a coherent state. For definitions see e.g. reference \cite{Tak1989}. 

We define two general localization measures in terms of the Husimi
quasi-probability distribution $H(q,p)$ in a phase space,
which can be computed for any quantal physical system. They are
normalized, $\int dqdp\, H(q,p) =1$.

The first localization measure is the effective volume $A$ on the classical
chaotic domain, covered by the Husimi function. $A$ is defined as
\begin{equation}
  A=\frac{e^{\left<I\right>}}{\Omega_c}
  \label{eqDegGenA}
\end{equation}
where
\begin{equation}
  I=-\int dq\,dp\,H(q,p)\log{H(q,p)}
  \label{eqDefGenI}
\end{equation}
is the information entropy, $\langle I\rangle$ is its average over the
large number of consecutive chaotic eigenstates and $\Omega_c$ is the phase
space volume of the chaotic domain on which $H(q,p)$ is defined.
Clearly, in the case of 
the uniform distribution $H(q,p)=1/\Omega_c$, the localization measure is $A=1$,
whilst in the case of the strongest localization (in a single Planck cell) 
$I=\log(2\pi\hbar)^f$, $A=(2\pi\hbar)^f/\Omega_c=1/N_C(E) \approx 0$, where
$f$ is the number of degrees of freedom, i.e. the dimension of the configuration
space, and $N_C(E)$ is the number of chaotic levels in the chaotic region 
at the energy $E$. In the semiclassical limit $2\pi\hbar \rightarrow 0$
this number is very large and thus $A\approx 0$.

For the definition of the second localization measure we first define a correlation
matrix

\begin{equation}
  C_{nm}=\frac{1}{Q_n\,Q_m}\int dq\,dp\,H_n(q,p)\,H_m(q,p)
  \label{eqDefGenC}
\end{equation}
where
\begin{equation}
  Q_n=\sqrt{\int dq\,dp\,H_n^2(q,p)}
  \label{eqDefGenNfact}
\end{equation}
are the normalization factors and
where $n$ and $m$ are just the eigenstate labels (quantum numbers).
It is clear that if two Husimi functions, $H^n$ and $H^m$
  do not "live" in the same part of
the phase space, their matrix element $C_{nm}$ is zero. This is possible,
if the corresponding eigenstates are from the different invariant domains,
or if the eigenstates are nonoverlapping on the chaotic region.
The correlation
matrix is therefore a useful object to study the clustering of eigenstates on the
chaotic domain. Thus we define the second measure of localization $C$
as an average of $C_{nm}$ over the sufficiently large number of consecutive
chaotic eigenstates
\begin{equation}
  C=\left<C_{nm}\right>.
  \label{eqDefGenCmeasure}
\end{equation}
Very interestingly and surprisingly, the numerical computations, 
explained in the next section, show that 
these two localization measures are linearly proportional and thus
equivalent.

\section{The model billiard system}

Billiards are nontrivial and generic dynamical systems in which it is
possible to compute a great number of high quality high lying energy
levels, using many elegant numerical techniques \cite{VebProRob2007}, especially 
the plane wave decomposition method \cite{VerSer1995}, all of them
used in \cite{BatRob2010}.
We choose the family of billiards introduced by Robnik
\cite{Rob1983,Rob1984}, 
defined as the quadratic conformal map $w=z+\lambda z^2$ of the unit 
circle $|z|=1$ from the complex $z$-plane onto the physical $w$-complex
plane, and study numerically a series of shapes at various $\lambda$. 
By varying the parameter $\lambda$ from $0$ to $1/2$ we see the 
transition from an integrable (circle) to the fully chaotic system
\cite{Markarian1993}, which is ergodic, mixing and K. 
In between the system is of the mixed-type with coexisting
regular and chaotic regions, a typical KAM-scenario, for 
$\lambda \ge 0.135$ having only one dominant chaotic region, so that
smaller chaotic regions can be neglected. It has been
already shown that the BRB distribution gives an
excellent description of the level spacing distribution \cite{BatRob2010}.
One example is shown in figure \ref{figPS}.
Moreover, very recently \cite{BatRob2013,BatManRob2013} 
it has been explicitly demonstrated in the case of $\lambda=0.15$,
with great accuracy and statistical significance ($\sim 590.000$ consecutive eigenstates), 
that the regular and chaotic eigenstates and the corresponding energy levels
can be separated, by means of Poincar\'e Husimi functions, clearly
yielding the Poisson statistics for the regular and Brody
level spacing distribution for the chaotic eigenstates.

In order to study the localization effects and their relationship to $\beta$, 
we have explored a series of billiards with various values 
$\lambda=$ $0.135$, $0.14$, $0.145$, $0.15$, $0.155$, $0.16$, $0.165$, $0.175$, 
$0.18$, $0.19$, $0.2$, $0.21$, $0.22$, $0.23$, $0.24$.
We have solved numerically the Helmholtz equation 
$\Delta \psi + k^2 \psi =0$, with the Dirichlet boundary
conditions $\psi=0$ for each $\lambda$. The size of the chaotic component
$\rho_c$ (the relative phase space volume of the chaotic region, 
not to be confused with the area of the chaotic region on the Poincar\'e
surface of section) and the degree of chaos increase monotonically with $\lambda$.
The ratio $\alpha=t_H/t_T$ of the Heisenberg time $t_H$ and the transport time $t_T$
is calculated as $\alpha=2k/N_T$, where $N_T$ is the characteristic
transport time in units of mean free flight time, i.e. it is the number
of collisions necessary for the global classical transport. The
dimensionless Heisenberg time is equal to $2k$.
For example, for $\lambda=0.15$ it turns out that $N_T\approx 10^5$.
If $\alpha < 1$, the quantum localization occurs. See \cite{BatRob2013}
for details.
The billiards at different $\lambda$ have different transport times $N_T$  
on the largest (dominant) chaotic component, and thus there is a different
degree of localization at fixed $k$. The details of the estimates of $N_T$
are given in the appendix A.
We have calculated  the localization measures 
for all billiards at two different $k$, $k\approx2000$ and $k\approx4000$,
and also the corresponding $\beta$.

We now define the localization measures in terms of the positive definite
Husimi function, which is interpreted as a probability distribution 
of a quantum state in the phase space, and is defined below.

The classical billiard dynamics is described completely by
the bounce map, using the Poincar\'e Birkhoff coordinates
$(q,p)$. Thus it is reasonable to use the Poincar\'e Husimi function which is a
quasi-probability distribution in the Poincar\'e Birkhoff phase space
\cite{Baecker2004,BatRob2013}, defined as
\begin{equation}
    H(q,p)=\left|\int_{\partial{\cal B}}c_{(q,p),k_{n}}(s)\,u_{n}(s)ds\right|^2,
    \label{eqDefHusimi}
\end{equation}
where
$    u_{n}(s)=\mathbf{n}(s).\nabla\psi_{n}(\mathbf{r}(s))$
is the normal derivative of the eigenfunction $\psi_{n}$ on the boundary,
with ${\bf n}(s)$ being the unit outward vector to the boundary $\partial{\cal B}$ 
at position $s$.
It is called also {\em the boundary function}, because it uniquely determines the
wave function at any point inside the billiard.  
For details see \cite{BatRob2013}. Here 
\begin{equation}
    c_{(q,p),k}=\sum_{m\in\mathbb{Z}}e^{ikp(s-q+mL)}\,e^{-k(s-q+mL)^2/2}
    \label{eqCohState}
\end{equation}
is the coherent state on the boundary (obviously periodized, thus satisfying the
periodic boundary condition in $s$), centered at $(q,p)$. 
There is indeed no loss of information with this representation as the boundary
function $u_n(s)$ gives the complete description of the eigenfunction in
the interior of the billiard ${\cal B}$.

\begin{figure}
  \centering
  \includegraphics[width=8.5cm]{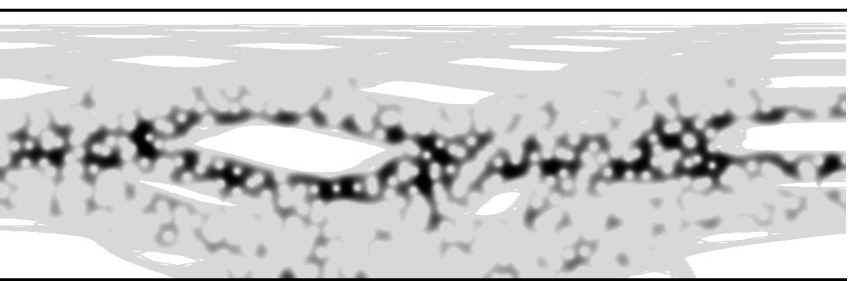}
  \includegraphics[width=8.5cm]{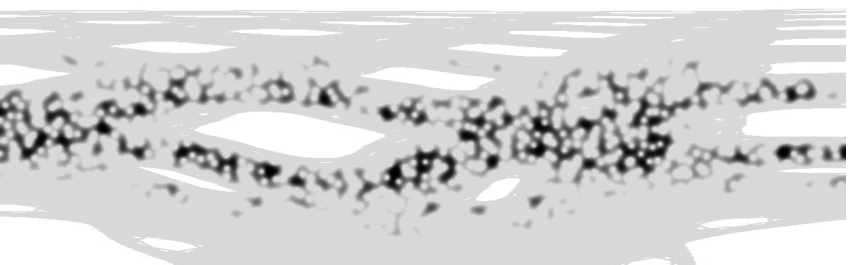}
  \caption{Poincar\'e Husimi functions for similar eigenstates of the $\lambda=0.15$
    billiard at $k\approx2000$
  (top) and $k\approx4000$ (bottom), showing the localization effect. 
The shaded area is the classical chaotic
domain, which would be completely covered by the extended states in a deep
semiclassical limit. Unshaded area represents the domains of regular
motion.}
  \label{figHusimiFun}
\end{figure}

$H(q,p)$ has been
calculated on the equidistant $400\times400$ grid
$(q_{i},p_{j})=\left(\Delta{q}/2+i\,\Delta{q},\Delta{p}/2+j\,\Delta{p}\right)$
where $\Delta{q}=L/800$ and $\Delta{p}=1/400$. The grid covers the quadrant
$q\in[0,L/2]$ and $p\in[0,1]$, which remains after the reduction of the
phase space due to the time reversal and the reflection symmetries.
The grid points are positioned at the centers of the square cells, of the area
$\Delta{q}\,\Delta{p}$. 
The integration method used to evaluate (\ref{eqDefHusimi}) is a simple trapeze
rule with the step $ds\propto\lambda_B/20$, where $\lambda_B=2\pi/k$ is the
de Broglie wavelength.  
Importantly, the values of the Poincar\'e Husimi function on the 
grid $H_{ij}=H(q_i,p_j)$ are rescaled such that their sum equals one (normalization of $H$).
Two examples of the localized chaotic 
eigenstates are shown in figure \ref{figHusimiFun}.

In order to separate the regular and chaotic eigenstates we have
introduced the overlap index $M$, following our previous work
\cite{BatRob2013,BatManRob2013}.
Each cell is ascribed the number $\gamma_{ij}$, which is either $+1$ if
it belongs to the chaotic region, or $-1$ if the region is regular. These
numbers are calculated classically by means of a very long chaotic orbit
(about $10^9$ collisions).
For a cell to be classified as chaotic only one visit of the chaotic
orbit is sufficient. The overlap index as the classification
measure is defined
as the sum $M=\sum_{ij}H_{ij}\gamma_{ij}$. Ideally, if $M=+1$, the level is
labeled chaotic  and if $M=-1$ the level is labeled regular. 
Using $M$ it is possible to extract chaotic states.

We express the localization measures in terms of the discretized Husimi
function.
For $A$ we write
\begin{equation}
  A = \frac{e^{\left<I\right>}}{N_c}
  \label{eqPovrsina}
\end{equation}
where 
\begin{equation}
  I = -\sum_{ij}H_{ij}\,\log{H_{ij}}
  \label{eqEntopija}
\end{equation}
and $N_c$ is a number of cells on the classical chaotic domain. 
Again, in the case of uniform dsitribution $H_{ij}=1/N_C$ the localization
measure is $A=1$, whilst in the case of the strongest localization
$I=0$, and $A=1/N_C \approx 0$. The mean $\left<I\right>$ is obtained by
averaging $I$ over a sufficiently large number of consecutive chaotic
eigenstates.

The correlation matrix equals
\begin{equation}
  C_{nm}=\frac{1}{Q_n\,Q_m}\sum_{ij}H^n_{ij}\,H^{m}_{ij},
  \label{eqKorelacija}
\end{equation}
where
\begin{equation}
  Q_n=\sqrt{\sum_{ij}(H^n_{ij})^2}
  \label{eqQ}
\end{equation}
is the normalizing factor.

\begin{figure}
\centering
\includegraphics[]{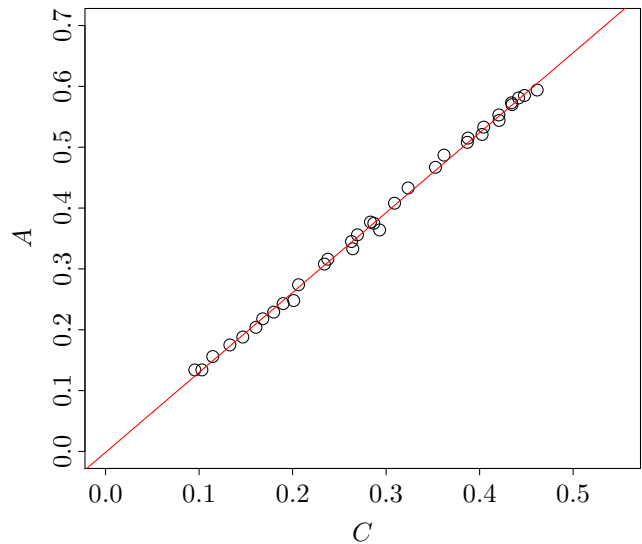}
\caption{Linear relation between the two entirely different localization
measures, namely the enrtopy measure $A$ and the correlation measure $C$,
calculated for several different billiards at $k\approx2000$ and
$k\approx4000$.}
\label{figCovVSCor}
\end{figure}

\begin{figure}
  \centering
  \includegraphics{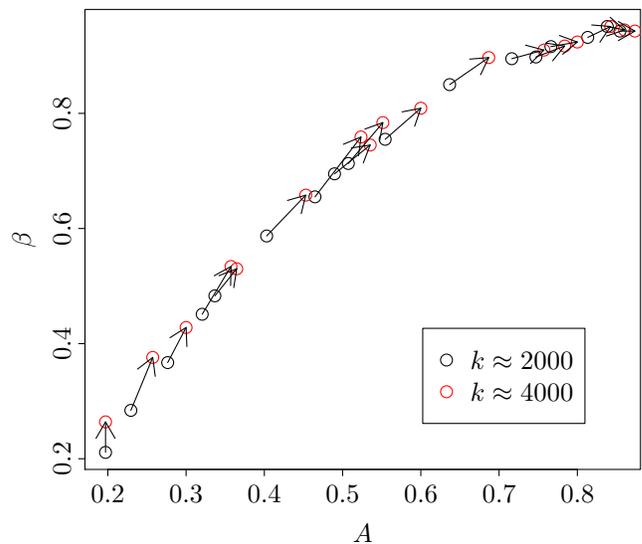}
  \caption{The central result of this letter: A clear functional relation between $\beta$
    and the localization measure $A$. 
    Arrows connect points corresponding to the same $\lambda$ at two different $k$. }
  \label{figBetVSA}
\end{figure}

For a good approximation of the localization measures $A$ and $C$ it was
sufficient to separate and extract about $1.500$ consecutive chaotic eigenstates.
It is very interesting and satisfying that the two
localization measures are linearly equivalent as shown in figure
\ref{figCovVSCor}. 
But to get a good estimate of $\beta$ we need much more levels, and the
separation of eigenstates is then technically too demanding.
We have instead calculated spectra on small intervals around $k\approx2000$ and
$k\approx4000$ taking not less than $100.000$ consecutive levels (no separation) 
and fitted their level spacing distribution with the BRB distribution with 
the $\beta$ as the only fit parameter, while using the fixed classically 
calculated parameter $\rho_r$. The dependence of $\beta$ on $A$ is shown 
in figure \ref{figBetVSA}. For aesthetic reasons we have rescaled the
measure $A\rightarrow A/A_\textrm{\small max}$ such that it goes from 0 to 1. 
The maximal value of $A$, $A_\textrm{\small max}=0.68$, was estimated
as $A_{\textrm{\small max}}=e^{I_\textrm{\tiny max}}/N_c$, where
$I_\textrm{\small max}$ is the 
maximum entropy of $1500$ consecutive states of the almost fully
chaotic $\lambda=0.25$ billiard. This is some kind of renormalization of
$A$, such that for fully chaotic systems the procedure always yields $A=1$.
Namely, in real chaotic eigenstates we never reach a perfectly uniform
distribution $H(q,p)$, since they always have some oscillatory structure.

We clearly see that there is a functional relationship between $A$ and $\beta$.
By increasing $k$ from $2000$ to $4000$ we increase the dimensionless Heisenberg 
time by factor $2$, therefore the degree of localization should decrease, meaning 
that $A$ must increase, but precisely in such a way, that the empirical points 
stay on the scaling curve, as it is observed and indicated by the arrows.

Unfortunately, it is too early to propose a semiempirical functional
description of the relationship we found in figure  \ref{figBetVSA}.
In the quantum kicked rotator it is just almost linear \cite{Izr1990,ManRob2013,BatManRob2013}. 
Also, there is a great lack in theoretical understanding of its 
physical origin, even in the case of (the long standing research on) 
the quantum kicked rotator, 
except for the intuitive idea, that energy spectral properties should be 
only a function of the degree of localization, because the localization
gradually decouples the energy eigenstates and levels, switching the linear
level repulsion $\beta=1$ (of fully extended chaotic states) to a power law
level repulsion with  exponent $\beta < 1$. The full physical 
explanation is open for the future.

\section{Conclusions}

Our main conclusion is that in autonomous Hamilton systems in quantum
chaos the spectral level repulsion parameter of the chaotic eigenstates
is functionally related to a localization measure. In our method we have defined two
localization measures, one in terms of the information entropy, and the
other one in terms of the correlation properties of Husimi functions.
Although different by definitions, we show in the case of the billiard
systems \cite{Rob1983,Rob1984}, working with the Poincar\'e Husimi
functions, that they are linearly proportional and thus equivalent. 
Our results are in complete analogy with the quantum kicked rotator.
Further theoretical work is in progress. Beyond the billiard systems,
there are many important applications in various physical systems,
like e.g. in hydrogen atom in strong magnetic field \cite{Rob1981,Rob1982,HRW1989,
WF1989,RWHG1994}, which is a paradigm of stationary quantum chaos,
or e.g. in microwave resonators, the experiments introduced by St\"ockmann
around 1990 and intensely further developed since then \cite{Stoe}.

\section{Acknowledgement}

This work was supported by the Slovenian Research Agency (ARRS).

\section*{Appendix A: Classical transport times}

Here we calculate the Heisenberg time and the classical
transport time for a chaotic billiard. According to the
leading order of the Weyl formula, which is in fact just the simple
Thomas-Fermi rule, we have for the number of levels $N(E)$ below
and up to the energy $E$ of a Hamiltonian $H({\bf q},{\bf p})$
\begin{equation} \label{A1}
N(E) = \frac{1}{(2\pi\hbar)^2} \int_{H({\bf q},{\bf p})\le E} d^2{\bf q}\;d^2{\bf p}.
\end{equation}
Since $H= {\bf p}^2/(2m)$, with constant zero potential energy
inside ${\cal B}$, where $m$ is the mass of the billiard
point particle, and $H$ is infinite on the boundary $\partial {\cal B}$,
we get at once
\begin{equation} \label{A2}
N(E) = \frac{2\pi {\cal A}mE}{(2\pi\hbar)^2}.
\end{equation}
Here ${\cal A}$ is the area of the billiard ${\cal B}$.
The density of levels is $\rho (E) = 1/(\Delta E) = 
dN(E)/dE = {\cal A}m/(2\pi\hbar^2)$
and thus the Heisenberg time is
\begin{equation}  \label{A3}
t_H = 2\pi\hbar \rho(E) = \frac{{\cal A} m}{\hbar}.
\end{equation}
The classical transport time is denoted by $t_T$, and in units of
the number of collisions $N_T$ can be written as 
\begin{equation} \label{A4}
t_T = \frac{\bar{l} N_T}{v} = \frac{ \bar{l}N_T}{\sqrt{2E/m}},
\end{equation}
where $\bar{l}$ is the mean free path of the billiard particle
and $v =\sqrt{2E/m}$ is its speed at the energy $E$. Thus for 
the ratio  $\alpha = t_H/t_T$ we get
\begin{equation} \label{A5}
\alpha = \frac{t_H}{t_T} = \frac{{\cal A} k}{N_T \bar{l}} 
\end{equation}
where $k= \sqrt{2mE/\hbar^2}$. Taking into account that
$\bar{l} \approx \pi {\cal A}/{\cal L}$ (this is so-called Santalo's
formula, see e.g. \cite{Santalo}), we have
\begin{equation} \label{A6}
\alpha = \frac{t_H}{t_T} = \frac{{\cal L}k}{\pi N_T},
\end{equation}
where ${\cal L}$ is the length of the perimeter $\partial {\cal B}$.
This is a general formula valid for any chaotic billiard.
In the case of Robnik billiards \cite{Rob1983,Rob1984} 
${\cal L}\approx 2\pi$ and we arrive at the final estimate
\begin{equation} \label{A7}
\alpha = \frac{2k}{N_T}.
\end{equation}
Thus the condition for the occurrence of dynamical localization
$\alpha \le 1$ is now expressed in the inequality

\begin{equation} \label{A8}
k \le \frac{N_T}{2}.
\end{equation}

\begin{figure}
\centering
\vspace{10pt}
\includegraphics{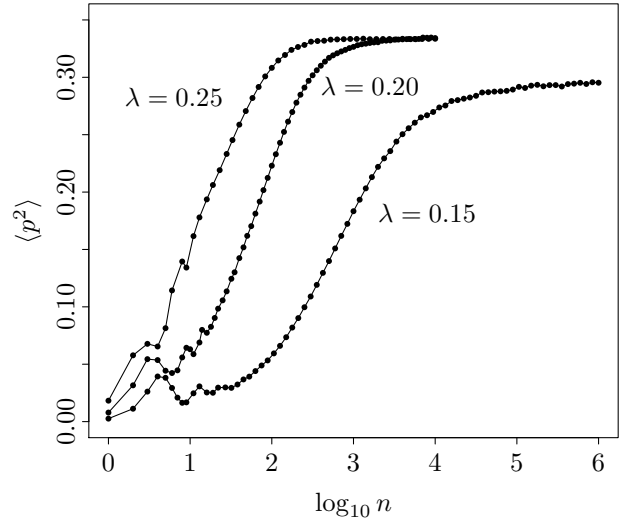}
\caption{We show the second moment $\langle p^2 \rangle$ averaged
over an ensemble of $10^6$ initial conditions  uniformly distributed
in the chaotic component on the interval $s\in [0,{\cal L}/2]$ and $p=0$ as a function of 
the decadic logarithm of  the 
number of collisions $n$. We see that the saturation value of 
$\langle p^2 \rangle$ is reached at about $N_T=10^5$ collisions for
$\lambda=0.15$, $N_T=10^3$ collisions for $\lambda=0.20$ and
$N_T=10^2$ for $\lambda=0.25$. For $\lambda=0.15$, 
according to the criterion (\ref{A8}) at
$k=2000$ and $k=4000$ we are still in the regime where the dynamical
localization is expected. On the other hand, for $\lambda=0.20,0.25$ we
expect extended states already at $k<2000$.}
\label{trans}
\end{figure}

Some examples of chaotic spreading of an inital ensemble
of orbits placed at $p=0$ on the chaotic region are shown 
in figure \ref{trans}, where we plot the mean square of the 
momentum as a function of the number of collisions, for
three different values of $\lambda$.

\bibliography{qChaos.bib}


\end{document}